\documentclass{pasj01}

\RequirePackage{mathpazo}
\RequirePackage[T1]{fontenc}

\input{pasjsup.sty}

\newcommand{\tc}[1]{#1}

\def\commenta{$^*$}
\def\commentb{$^\dagger$}
\def\commentc{$^\ddagger$}
\def\commentd{$^\S$}
\def\commente{$^\|$}

\newcounter{author}
\def\authorcount#1#2{\refstepcounter{author}\label{#1}
                     \altaffiltext{\ref{#1}}{#2}}

\begin{document}
\SetRunningHead{K. Isogai et al.}{NSV 1440}

\Received{201X/XX/XX}%{yyyy/mm/dd}
\Accepted{201X/XX/XX}%{yyyy/mm/dd}

\title{NSV 1440: First WZ Sge-type Object in AM CVn stars and candidates}
\author{Keisuke~\textsc{Isogai},\altaffilmark{\ref{affil:Kyoto}*}
  Taichi~\textsc{Kato},\altaffilmark{\ref{affil:Kyoto}}
  Berto~\textsc{Monard},\altaffilmark{\ref{affil:Monard}}
  Franz-Josef~\textsc{Hambsch},\altaffilmark{\ref{affil:GEOS}}$^,$\altaffilmark{\ref{affil:BAV}}$^,$\altaffilmark{\ref{affil:Hambsch}}
  Gordon~\textsc{Myers},\altaffilmark{\ref{affil:Myers}}
  Peter~\textsc{Starr},\altaffilmark{\ref{affil:Starr}}
  Lewis~M.~\textsc{Cook},\altaffilmark{\ref{affil:LewCook}}
  Daisaku~\textsc{Nogami}\altaffilmark{\ref{affil:Kyoto}}
}

\authorcount{affil:Kyoto}{
     Department of Astronomy, Kyoto University, Kyoto 606-8502, Japan}
\email{$^*$isogai@kusastro.kyoto-u.ac.jp}

\authorcount{affil:Monard}{
Bronberg and Kleinkaroo Observatories, Center for Backyard Astrophysics Kleinkaroo,
Sint Helena 1B, PO Box 281, Calitzdorp 6660, South Africa}

\authorcount{affil:GEOS}{
     Groupe Europ\'een d'Observations Stellaires (GEOS),
     23 Parc de Levesville, 28300 Bailleau l'Ev\^eque, France}

\authorcount{affil:BAV}{
     Bundesdeutsche Arbeitsgemeinschaft f\"ur Ver\"anderliche Sterne
     (BAV), Munsterdamm 90, 12169 Berlin, Germany}

\authorcount{affil:Hambsch}{
     Vereniging Voor Sterrenkunde (VVS), Oude Bleken 12, 2400 Mol, Belgium}

\authorcount{affil:Myers}{
     Center for Backyard Astrophysics San Mateo, 5 inverness Way,
     Hillsborough, CA 94010, USA}

\authorcount{affil:Starr}{
     Warrumbungle Observatory, Tenby, 841 Timor Rd,
     Coonabarabran NSW 2357, Australia}

\authorcount{affil:LewCook}{
     Center for Backyard Astrophysics Concord, 1730 Helix Ct. Concord,
     California 94518, USA}

%%% end:list of authors

\KeyWords{accretion, accretion disks
          --- stars: novae, cataclysmic variables
          --- stars: dwarf novae
          --- stars: individual (NSV 1440)
         }

\maketitle

\begin{abstract}
In 2015 and 2017, the AM CVn \tc{candidate} NSV 1440 showed superoutbursts having
the characteristic features of WZ Sge-type dwarf novae (DNe). 
By analogy with hydrogen-rich cataclysmic variables (CVs),
we can interpret these outbursts as ``double superoutbursts''
which are composed of the first superoutburst with early superhumps
and the second superoutburst with ordinary superhumps.
The object also showed multiple rebrightenings after the main superoutbursts.
Early superhumps had been never observed in AM CVn stars \tc{and candidates},
thus NSV 1440 is the first \tc{confirmed} WZ Sge-type AM CVn \tc{candidate}.
We obtained the early superhump period of 0.0252329(49) d and
the growing (stage A) superhumps period of 0.025679(20) d from the 2015 superoutburst.
We regarded the early superhump period as the orbital one.
By using these periods we estimated the mass ratio $q = $0.045(2).
This value suggests that \tc{NSV 1440 is indeed an AM CVn star and that the secondary is a semi-degenerate star}.
\end{abstract}

\section{Introduction}
AM CVn stars are a subclass of cataclysmic variables (CVs),
which are close binary systems composed of a white dwarf (WD) primary
and a mass-transferring secondary.
Their secondary stars are helium stars or helium WDs.
They are characterized by absences of hydrogen lines in their spectra
and their ultra-short orbital periods of 5--65 min
(for reviews of AM CVn stars, see \cite{nel05amcvnreview,sol10amcvnreview}).

\tc{Outbursts} in AM CVn stars are theoretically studied by \citet{tsu97amcvn}
and are basically understood by analogy with the thermal instability model in H-rich dwarf novae (DNe).
However, the outburst behaviors of AM CVn stars are complicated, e.g. dips, rebrightenings and so on.
\citet{kot12amcvnoutburst} argued that variation of the mass-transfer rate,
may be caused by irradiation of the secondary, is necessary for
reproducing their light curves on the basis of their model calculation.
\citet{war95amcvn} and \citet{war15amcvnmemsai} interpreted
the typical outbursting AM CVn stars
as VY Scl-type objects, which show brightness variations owing to the change of the mass-transfer rate.
More detailed observational studies are required to understand the outburst mechanism
and the stability of the mass-transfer rate.

Some AM CVn stars show not only normal outbursts
but also superoutbursts with superhumps \citep{pat93amcvn,war95amcvn}.
Superhumps are small-amplitude modulations whose period $P_{\rm SH}$ is
a few percent longer than the orbital period $P_{\rm orb}$.
Superoutbursts and superhumps are characteristic phenomena of SU UMa-type DNe
and are explained by the thermal-tidal instability (TTI) model \citep{osa89suuma}.
When an outer disk reaches the 3:1 resonance radius,
the disk becomes eccentric and begin to show periodic modulations,
i.e. superhumps \citep{whi88tidal,lub91SHa,lub91SHb,hir90SHexcess}.
\citet{kat14j0902} and \citet{iso16crboo} confirmed that
the period variations of superhumps in AM CVn stars are consistent with those in H-rich DNe,
and proposed that the superoutbursts in AM CVn stars are also interpreted by the TTI model.

AM CVn stars typically have an extreme low mass \tc{secondary (cf. \cite{nel01amcvnpopulationsynthesis}).
We know that $P_{\rm orb}$ of outbursting AM CVn stars are typically longer than 1300 s $\sim$ 0.015 d \citep{sol10amcvnreview},
and that $M_1$ of $0.65 M_{\odot}$ is often used (e.g. \cite{bil06amcvn}).
From these values and the theoretical evolutionary tracks (see section \ref{sec:q} and equation \ref{eq:fulldege} and \ref{eq:semidege}),
we can approximately estimate that $q$ of typical outbursting AM CVn stars are less than 0.1.
}
Because disks in CVs with $q <$ 0.25--0.30 can reach the 3:1 resonance radius,
many outbursting AM CVn stars can show superhumps.

\tc{
Many hydrogen-rich DNe with $q < 0.09$ are known as WZ Sge-type DNe, which are a subclass of SU UMa-type DNe.
It is known that WZ Sge-type DNe show longer and larger superoutburst in comparison with SU UMa-type DNe
and show few normal outbursts (for a review of WZ Sge-type DNe, see \cite{kat15wzsge}).
The analogy between WZ Sge-type DNe and some AM CVn stars has sometimes been discussed.
For instance, \citet{nog04v406hya} pointed out that the outburst behavior of V406 Hya resembles with that of the WZ Sge-type star EG Cnc
in that the object showed the multiple rebrightenings after the main superoutburst.
\citet{lev15amcvn} proposed that the long $P_{\rm orb}$ systems have low mass-transfer rates and will show rare and large outbursts.
Thus, they indicated that such objects may be WZ Sge-type AM CVn stars.
To date, ``WZ Sge-type'' in AM CVn stars has basically meant that an object 
shows multiple rebrightenings or that an object shows rare and larger-scale superoutbursts.
}

\tc{
The larger-scale WZ Sge-type superoutbursts are explained by the presence of the 2:1 resonance.
If a system has extreme low mass ratio and enough mass is accumulated in the disk,
the outer edge of the disk can reach the 2:1 resonance radius beyond the 3:1 one \citep{osa02wzsgehump}.
The two-armed dissipation pattern in the disk is caused by the 2:1 resonance, 
then the early superhumps begin to grow \citep{lin79lowqdisk}.
Because the 2:1 resonance suppresses the 3:1 resonance \citep{lub91SHa,osa03DNoutburst},
ordinary superhumps begin to grow after the end of the early superhump phase.
It is widely known that early superhumps have double-wave profiles
and the periods are close to $P_{\rm orb}$ \citep{kat02wzsgeESH}.
For instance, \citet{ish02wzsgeletter} have confirmed that the early superhump period of AL Com is 0.05 $\%$ shorter than $P_{\rm orb}$.
\citet{pat02wzsge} also proposed that the signal of the early superhumps in WZ Sagittae is ``essentially consistent with orbital frequency''.
It is considered that a vertical extended disk originates such modulations.
Thus, early superhumps are only observed in high-inclination systems.
On the other hands, low-inclination systems show a long plateau phase with no superhumps 
which is brighter than the superoutburst plateau with ordinary superhumps.
Such a plateau phase is also regarded as a kind of early superhump phases.
Because WZ Sge-type superoutbursts are essentially different from SU UMa-type ones,
WZ Sge-type DNe are defined by the presence of the early superhump phase according to the modern criteria \citep{kat15wzsge}.
For these reasons, WZ Sge-type superoutbursts are brighter and longer than SU UMa-type ones.
}

\tc{As indicated in \citet{lev15amcvn},} many AM CVn stars could show WZ Sge-type superoutbursts
since they have low $q$ and low mass-transfer rate,
especially in long period systems.
\tc{Actually, some objects have shown WZ Sge-like light curves.
However, the reliable evidence of WZ Sge-type superoutbursts, 
namely the early superhump phase, has never been observed in AM CVn stars and candidates.
Because we can estimate $P_{\rm orb}$ from the early superhump period,
WZ Sge-type superoutbursts are not just large and rare superoutbursts but the important messengers of the binary parameters.
Furthermore, we can also estimate the mass ratios from intensive time-series observations of WZ Sge-type superoutbursts and 
evaluate the evolutionary path as will be discussed in section \ref{sec:q}.
}

In this paper, we report on our time-series observations
of the 2015 and 2017 outbursts of the AM CVn \tc{candidate} NSV 1440.
\tc{Although there is no spectroscopic confirmation, 
the short period superhump and outburst behavior suggest that the object is an AM CVn star.
Actually, the object is often treated as an AM CVn star, e.g. \citet{ram18amcvngaia}.}
The object showed the first WZ Sge-type superoutbursts in AM CVn \tc{stars and candidates}.
This fact implies that we can understand outbursts in AM CVn stars by analogy with hydrogen-rich DNe.

\section{NSV 1440}\label{sec:nsv1440}
NSV 1440 was a variable star candidate listed in the New Catalogue of Suspected Variable Stars (NSV, \cite{NSV})
with the brightness range from 12.6 up to 15.0 mag.
The object is also known by the names BV 1025, 
ASASSN-15sz and GALEX J035517.7-822612.
The coordinates of the object are RA = 03:55:17.83 and Dec = -82:26:11.5 at J2000.
The quiescent magnitudes in {\it Gaia} Data Release 2 are $G=18.5126(72)$,
$BP=18.4139(331)$ and $RP=18.4239(684)$ \citep{Gaia2016,Gaia2018,GaiaDR2photmetry}.
By using these values and table A.2 in \citet{GaiaDR2photometry2},
we can estimate the quiescent $V$ mag of 18.53(5).
The object has a GALEX counterpart with near-UV (NUV) and far-UV (FUV)
magnitudes of 18.305(57) and 18.286(96) \citep{GALEX}.
Two historical outbursts were recorded in the All Sky Automated Survey-3 (ASAS-3, \cite{ASAS3}),
\tc{cf. the light curve of ASAS-3 in figure E2.}
The 2003 outburst was detected at 13.544 mag on BJD 2452929.740037,
and the 2005 one was detected at 13.492 mag on BJD 2453669.797309.

The 2015 outburst was detected at $V=$ 13.8 on November 20 (BJD 2457339.62)
by the All-Sky Automated Survey for Supernovae (ASAS-SN) \citep{ASASSN}.
The 2017 outburst was detected at a visual magnitude of 13.0
on August 21 (BJD 2457987.27) by R. Stubbings (vsnet-alert 21352\footnote{
VSNET-alert archive is available at\\
<http://ooruri.kusastro.kyoto-u.ac.jp/pipermail/vsnet-alert/>
}).
After these detections, we performed the observation campaigns.

\begin{figure*}[htb]
\begin{center}
    \FigureFile(160mm,50mm){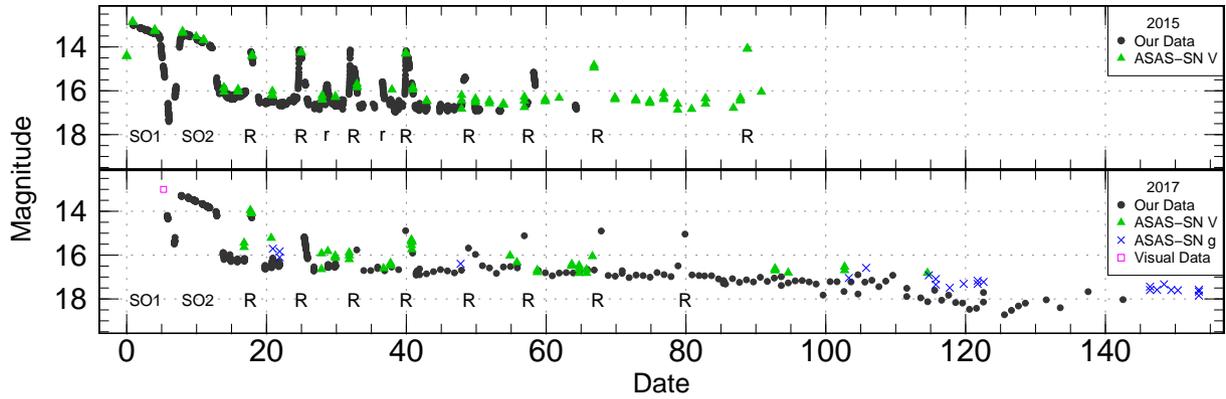}
\end{center}
\caption{Overall light curves of NSV 1440 in the 2015 and 2017 outbursts.
The observations were binned to 0.01 d.
Circles, triangles, crosses and a square represent our observations, the ASAS-SN $V$-band data,
the ASAS-SN $g$-band data and a visual observation by R. Stubbings, respectively.
The dates of the 2015 and 2017 outbursts are defined to be BJD $-$ 2457346.751 and BJD $-$ 2457982.000, respectively.
}
\label{fig:lc}
\end{figure*}

\section{Observation and Analysis}\label{sec:obs}
Our time-series observations are summarized in table E1 and E2\footnote{
Figure E1--E2 and tables E1--E4 are available online as the supplementary data for this article.
}.
\tc{The typical exposure time is 30--120 sec.}
The data were acquired by time-series unfiltered CCD photometry 
using 30--40 cm class telescopes by the VSNET Collaboration \citep{VSNET}.
The times of the observations were corrected to Barycentric Julian Date (BJD).
We adjusted the zero-point of each observer to the data of Franz-Josef Hambsch.

We used the phase dispersion minimization (PDM) method 
for analyzing the superhump periods.
We estimated $1\sigma$ errors by using the methods
in \citet{fer89error} and \citet{Pdot2}.
Before our period analyses, we subtracted the global trend of the light curve
which was calculated using locally-weighted polynomial regression (LOWESS, \cite{LOWESS}).
We used $O-C$ diagrams which are sensitive to subtle variations of the superhump period.
The times of superhump maxima, which are used to draw the $O-C$ diagrams
and are listed in table E3 and E4\footnotemark[2], were determined 
by the same method as described in \citet{Pdot}.
\tc{The number in the parentheses after each value represents $1\sigma$ error, e.g. $0.12(3)$ means $0.12 \pm 0.03$.}

\section{Result}\label{sec:result}
\subsection{Overall Light Curve}\label{sec:lc}
Figure \ref{fig:lc} shows the overall light curves of the 2015 and 2017 outbursts.
We also added the $V$ and $g$-band data obtained by the ASAS-SN Sky Patrol \citep{ASASSN,koc17ASASSNLC}.
We should note that the ASAS-SN data around 16.5 mag might include systematic errors
due to their limiting magnitudes.
The horizontal axis ``Date'' of the 2015 and 2017 outbursts are defined
to be BJD $-$ 2457346.751 and BJD $-$ 2457982.000, respectively.
Each light curve shows two superoutbursts (double superoutburst) and rebrightenings.
We respectively marked the superoutbursts, rebrightenings and 
``small rebrightenings'' with the labels ``SO'', ``R'' and ``r''.
The overall light curves are roughly consistent with each other.

According to the 2015 light curve, the maximum brightness value of
the first superoutburst (SO1) is V=12.85 on Date $=$ 0.85.
The SO1 lasted for about 4 d and the object rapidly faded to 18.0 mag.
Then the object brightened again and reached
the maximum of the second superoutburst (SO2) on Date $=$ 7.6.
The duration of the SO2 is about 5 d.
Unfortunately, there is no data of the plateau of the SO1 in 2017 due to the observational gap,
hence we cannot exclude the possibility that the SO1 in 2017 is not a superoutburst but a precursor outburst.
After the end of the SO2, the object showed a slow decline with multiple rebrightenings.
In 2017, the object returned to
the quiescent magnitude of V $=$ 18.53 around Date $=$ 120.

\begin{figure}[htb]
\begin{center}
    \FigureFile(80mm,50mm){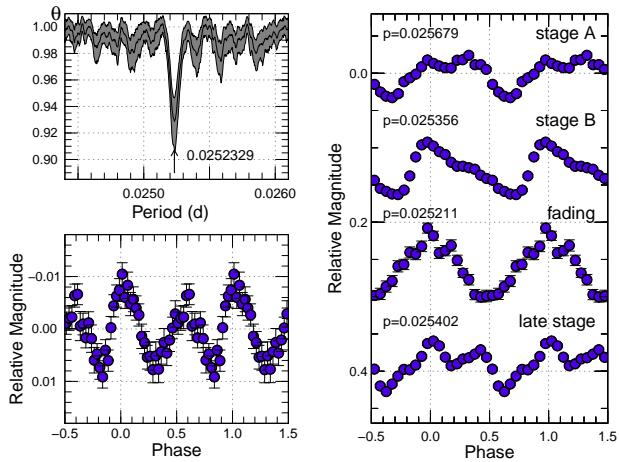}
\end{center}
\caption{Left: result of the period analyses of early superhumps in 2015.
(Upper): $\theta$ diagram of our PDM analysis of early superhumps.
The area of gray scale means 90\% confidence intervals
in the resultant $\theta$ statistics.
(Lower): phase-averaged profile of early superhumps.
Right: phase-averaged profiles of the modulations
in stage A, B, the fading stage and the late stage in 2015.
} 
\label{fig:pdm}
\end{figure}

\begin{figure}[htb]
\begin{center}
    \FigureFile(80mm,50mm){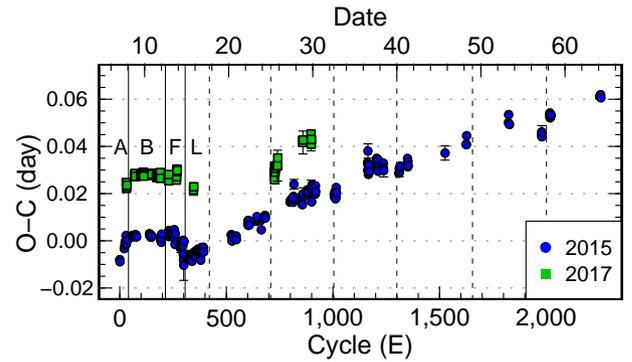}
\end{center}
\caption{$O-C$ diagrams of superhump maxima of NSV 1440 in 2015 (filled circle) and in 2017 (filled rectangle).
Ephemerises of BJD $=0.02535754750E + 2457353.794513$ (2015) and $+ 2457996.842150$ (2017) were used to draw this figure.
The vertical solid and dashed lines represent the timings of the superhump-stage transitions and the rebrightenings, respectively.
The horizontal axes show the cycle numbers and Date (same as figure \ref{fig:lc}).
The labels ``A'', ``B'', ``F'' and ``L'' mean stage A, B, the fading stage and the late stage, respectively.} 
\label{fig:oc}
\end{figure}

\subsection{Early superhumps}
During the SO1, we succeeded in detecting double-wave modulations with a constant period of 0.0252329(49) d.
The result of the period analysis and the averaged profile are
shown in the left panels of figure \ref{fig:pdm}.
\tc{Because the SO1 is brighter than the SO2 which shows the ordinary superhumps,
the disk should expand beyond the 3:1 resonance radius and reach the 2:1 one.
As explained in introduction, the 2:1 resonance suppresses the 3:1 resonance \citep{lub91SHa,osa03DNoutburst}.
If the disk did not reach the 2:1 resonance radius, the object should show ordinary superhumps during the SO1.
As explained in next section, the growing ordinary superhumps were detected in the rising part of the SO2,
thus we propose that the modulations in the SO1 are early superhumps.
}
The constant period and the characteristic double-wave profile
also suggest that they are early superhumps.
Thus, we can use the period of the early superhump as $P_{\rm orb}$.

\subsection{Ordinary superhumps}
After the end of the SO1, ordinary superhumps began to grow. 
We drew the $O-C$ diagrams of superhump maxima (figure \ref{fig:oc}).
For the classification and the interpretation of the superhump stages, see \cite{Pdot}  and \cite{kat13qfromstageA}.
In WZ Sge-type DNe, ``late-stage superhumps'' are usually observed
instead of stage C ones after the main superoutburst \citep{kat15wzsge}.
$P_{\rm SH}$ in the late stage is shorter than in stage A but longer than in stage B.

We can distinguish between stage A, B and late-stage superhumps from the $O-C$ diagrams.
Around the rapid fading from the SO1 (marked ``F'' in figure \ref{fig:oc}),
the object showed short-period modulations (hereafter we call this term ``fading stage'').
We will discuss the modulations in the fading stage in section \ref{sec:fadingstage}.
We summarized the estimated periods and the Dates used for our analysis in table \ref{tab:period}.
The periods in 2017 \tc{have significantly} larger errors due to the lacking data,
thus we will basically use the values in 2015.
The right panel of figure \ref{fig:pdm} shows phase-averaged profiles in 2015.

\begin{table}[htb]
    \caption{List of the estimated periods.}
  \label{tab:period}
\begin{center}
  \begin{tabular}{lll} \hline
    & Period (day)  &  Date \\ \hline\hline
\multicolumn{3}{c}{2015}\\ \hline
    Early superhump\commenta & 0.0252329(49) & 0.891--4.769  \\ 
    Stage A & 0.025679(20) & 6.872--8.082  \\ 
    Stage B & 0.025356(2) & 8.522--12.082 \\ 
    Fading stage & 0.025211(16) & 12.177--14.786 \\ 
    Late stage & 0.0254019(4)  & 14.850--64.348 \\ \hline
\multicolumn{3}{c}{2017} \\ \hline
    Stage A &  0.02562(21) & 7.790--7.899 \\ 
    Stage B & 0.0253478(17) & 8.787--11.897 \\ 
    Fading stage &  0.02531(17) & 12.776--12.903  \\ 
    Late stage & 0.0253926(7) & 14.770--29.892 \\ \hline
    \multicolumn{3}{l}{\commenta\hspace{0.5mm} We use the early superhump period as $P_{\rm orb}$.} \\ 
    \end{tabular}
\end{center}
\end{table}

\subsection{Rebrightenings}\label{sec:reb}
Each light curve shows eight rebrightenings.
The timings of the rebrightenings are almost the same except for the last one.
The duration and the amplitudes are 1.5 d and 2.1--2.6 mag, respectively.
Only in the 2015 outburst, the object showed two ``short rebrightenings''
whose duration and amplitudes are respectively $\sim$ 0.6 d and $\sim$ 1 mag
(you can see the enlarged light curve of the rebrightenings in figure E1\footnotemark[2]).

We extracted the linear rising/fading part of the rebrightenings in 2015
and evaluated the rising/fading rates.
The averaged rising/fading rates in the normal rebrightenings are
approximately -15(2) mag/d and 2.2(2) mag/d, respectively.
Such a rapid rising suggests that the normal rebrightenings are outside-in outbursts.
In contrast, the rising/fading rates in the short rebrightenings are about -6 mag/d and 4.7 mag/d, respectively.
The slow rising implies that the short rebrightenings are inside-out outbursts and arose only in the inner part of the disk.

\section{Discussion}\label{sec:discuss}
\subsection{WZ Sge-type superoutburst}\label{sec:ds}
As mentioned in the introduction, many AM CVn stars have low $q$
and can potentially cause a WZ Sge-type superoutburst.
A part of H-rich CVs with extreme low $q < 0.06$ show
double superoutbursts which are composed of the first superoutburst with early superhumps
and the second superoutburst with ordinary superhumps \citep{kat15wzsge}.
Because the growth time of the 3:1 resonance is proportional to $q^2$ \citep{lub91SHa},
the systems having low $q$ cannot maintain the superoutburst just after the disappears of the early superhumps.
When the ordinary superhump sufficiently develops,
the object undergoes the second superoutburst \citep{kim16asassn15jd}.
The outburst behaviors of NSV 1440 are in agreement with this interpretation.
Although we don't know the true $P_{\rm orb}$ of NSV 1440, we can interpret the modulations in the SO1
as early superhumps on the basis of the outburst morphology.
Thus NSV 1440 is the first promising WZ Sge-type \tc{DN in AM CVn stars and candidates}.

WZ Sge-type superoutbursts in AM CVn stars could be double superoutbursts because of their extreme low $q$.
\citet{lev15amcvn} investigated the long-term light curves of many AM CVn stars
and confirmed WZ Sge-type DNe-like light curves.
The light curve of SDSS J172102.48+273301.2 resembles
those of NSV 1440 closely (figure 10 in \cite{lev15amcvn}).
There is however no time-resolved data and we cannot confirm the presence of early superhumps.
SDSS J090221.35+381941.9 reported in \citet{kat14j0902} showed
a precursor outburst one week before the SU UMa-type superoutburst.
The precursor may have been a superoutburst with early superhumps,
but they missed the overall profile of the precursor.
Recently, SDSS J141118.31+481257.6 and ASASSN-18rg also showed
double superoutburst-like phenomena (Isogai et al. in preparation).

\subsection{Orbital-period modulations in the fading stage}\label{sec:fadingstage}
The periods of the modulations in the fading stage are
close to $P_{\rm orb}$ (table \ref{tab:period}).
GW Lib, a typical H-rich WZ Sge-type DN, also showed
such orbital-period modulations in the fading stage (see figure 33 in \cite{Pdot}).
It is known that the pressure effect in the disk shortens $P_{\rm SH}$ \citep{lub92SH}.
If the pressure effect is amplified in the fading stage,
$P_{\rm SH}$ might match with $P_{\rm orb}$.
The right panel of figure \ref{fig:pdm} shows the orbital-period modulations have a sine wave shape.
The period and the profile may give us the impression that the bright spot was brightened in the fading stage,
namely the mass transfer was enhanced.
However, it is difficult to understand the reason why
such orbital modulations become visible only in the fading stage.

\subsection{Mass ratio and evolutionary channel}\label{sec:q}
Three evolutionary channels (WD, helium-star and evolved-CV channels) 
have been proposed to form AM CVn stars,
but the contribution of each channel is poorly understood.
The secondaries of WD channel systems are fully-degenerate WDs.
Whereas, those of helium-star and evolved-CV channel systems are
initially semi-degenerate stars and gradually evolve into fully-degenerate ones (cf. \cite{del07amcvn}).
Thus, secondary masses help to reveal their evolutionary channels.

We can also use mass ratio $q$, which is easier to obtain than donor mass.
\tc{
The empirical relation between $q$, $P_{\rm orb}$ and $P_{\rm SH}$ of hydrogen-rich DNe has been widely known, e.g. \citet{pat05SH}.
However, \citet{roe06amcvn} confirmed that $q$ from the empirical law is significantly different from their spectroscopic measurement.
Furthermore, \citet{pea07amcvnSH} indicated that superhump periods are affected by the pressure effect in the disk
and computed the pressure effect of AM CVn stars.
According to his formulation, the pressure effect also depends on the mass-radius relation of the secondary.
Because the mass-radius relation of AM CVn stars differs with the evolutionary scenarios,
\citet{pea07amcvnSH} concluded that we should not use the empirical law for AM CVn stars.
\citet{osa13v344lyrv1504cyg} interpreted that $P_{\rm SH}$ of the growing (stage A) superhump
corresponds to the dynamical precession rate at the 3:1 resonance radius
based on the Kepler's complete light curve.
\citet{kat13qfromstageA} investigated $P_{\rm SH}$ and $q$ of hydrogen-rich CVs
and have established the $q$ estimation method in a purely dynamical way.
They confirmed that $q$ from stage A are in good agreement with $q$ from the eclipse measurement.
Because this method does not depend on the disk composition and the secondary mass-radius relation,
we can apply to other DN cousins.
In fact, Ohnishi et al. (submitted) estimated $q$ of the metal-poor (population II) system OV Boo 
by using $P_{\rm SH}$ in stage A and the early superhump period which is regarded as $P_{\rm orb}$.
They succeeded in confirming that $q$ from the stage A method is consistent with $q$ from the eclipse measurement.
The metal abundance significantly affects the mass-radius relation \citep{ste97CVpopulationII} 
and the viscosity in the disk \citep{poj86scurve}.
Therefore, this result strongly suggests that we can also apply the stage A method to AM CVn stars.
However, we need to confirm the orbital period and mass ratio via spectroscopies or eclipse observations.
}
\citet{kat13qfromstageA} proposed that $q$ can be estimated by using
the fractional superhump excess $\varepsilon^* \equiv 1-P_{\rm orb}/P_{\rm SH}$ in stage A.
On the basis of the theoretical equations in \citet{kat13qfromstageA} and \tc{$\varepsilon^* = 0.0174(8)$,
we obtained $q = 0.045(2)$}.

\begin{figure}[htb]
\begin{center}
    \FigureFile(80mm,50mm){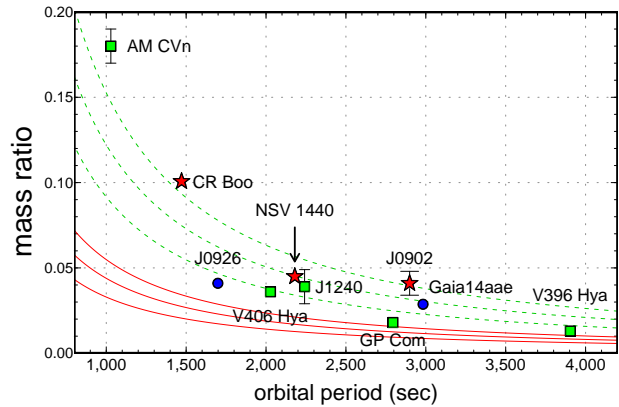}
\end{center}
\caption{Relation between $q$ and $P_{\rm{orb}}$ with various primary masses.
The dashed curve indicates semi-degenerate secondaries,
and the solid curve indicates fully-degenerate secondaries.
From top to bottom, three lines represent the $q$-$P_{\rm{orb}}$ relation, assuming 
$M_1 =$ 0.60, 0.75, and 1.00 $M_{\odot}$, respectively.
The filled stars represent the measurements from stage A superhump 
(CR Boo: \cite{iso16crboo}, J0902 = SDSS J090221.35+381941.9: \cite{kat14j0902}).
The filled squares represent the measurements from Doppler tomography \citep{DopplerTomography}
(AM CVn: \cite{roe06amcvn}, V406 Hya: \cite{roe06v406hya},
J1240 = SDSS J124058.03-015919: \cite{roe05j1240}, GP Com: \cite{mar99gpcom}, V396 Hya: \cite{sol10amcvnreview}).
The filled circle represents the measurement from eclipse observations
(J0926 = SDSS J092620.42+034542.3: \cite{cop11j0926}, Gaia14aae: \cite{gre18gaia14aae}).
} 
\label{fig:evol}
\end{figure}

\tc{
\citet{arm12cperi} derived the following evolutionary tracks from the Kepler's third law,
the secondary's Roche lobe-filling condition \citep{fau72amcvn},
the mass-radius relation of fully-degenerate stars \citep{zap69massradius}
and that of semi-degenerate stars \citep{sav86periodminimum}:
\begin{equation}\label{eq:fulldege}
M_2 = 1.43\times10^{-4}P_{\rm orb}^{-1.22} \ \ \ \ \ {\rm for\ the \ fully\mathchar`-degenerate\ secondary,}
\end{equation}
\begin{equation}\label{eq:semidege}
M_2 = 3.18\times10^{-4}P_{\rm orb}^{-1.27} \ \ \ \ \ {\rm for\ the \ semi\mathchar`-degenerate\ secondary,}
\end{equation}
}
Figure \ref{fig:evol} shows the above evolutionary tracks on the $P_{\rm orb}$-$q$ plane.
The dashed and solid curves respectively mean semi and fully-degenerate secondaries \tc{assuming $M_1$}.
The value of NSV 1440 suggests that the object has a semi-degenerate secondary,
and hence the object is \tc{an AM CVn star in} a helium-star or evolved-CV channel.

\subsection{Period variations of superhumps}\label{sec:pdot}
From the $O-C$ diagrams, we estimated the period derivative $P_{\rm dot} = \dot{P}/P$ of stage B
to be $-1.4(5) \times 10^{-5}$ in 2015 and $-1.5(7) \times 10^{-5}$ in 2017.
In hydrogen-rich CVs, it is empirically known that the $P_{\rm dot}$ depends on $P_{\rm orb}$ and $q$.
Especially, $P_{\rm dot}$ of WZ Sge-type DNe have good correlation with $q$.
\citet{kat15wzsge} derived the empirical relation $q=0.0043(9)P_{\rm dot}\times 10^5 + 0.060(5)$ for H-rich WZ Sge-type DNe.
Then, we obtained $q =$ 0.054(6) in 2015 and 0.053(6) in 2017 which
are larger than $q=0.045(2)$ estimated in section \ref{sec:q}.
We should modify the relation for AM CVn stars.

The pressure effect in the disks in He-rich CVs appears to be larger than that in H-rich CVs
because of the higher ionization temperature \citep{pea07amcvnSH}.
\citet{kat14j0902} argued that the period variation between stage A and B reflects the pressure effect.
For this purpose, we calculated $\varepsilon^*({\rm stage\ A})-\varepsilon^*({\rm stage\ B})=0.0125(8)$.
This value is in agreement with those in H-rich CVs of 0.010-0.015 \citep{Pdot} against the expectation.
The reason may be that the pressure effect also depends on other parameters, e.g. $q$ (see equation 24 in \cite{pea07amcvnSH}).

\subsection{Supercycle}\label{sec:supercycle}
\tc{As mentioned in section \ref{sec:nsv1440}, the four outbursts have been detected: 
$V=13.544$ in 2003, $V=13.492$ in 2005, $V=13.8$ in 2015, $V=13.0$ in 2017.
The complete light curves of ASAS-3 and ASAS-SN are shown in figure E2.
Although there is no time-resolved data in 2003 and 2005, their outburst maxima suggest that they are superoutbursts.
As you can see in figure \ref{fig:lc}, all superoutbursts are brighter than 14.0 mag, 
and all rebrightenings (normal outbursts) are fainter than 14.0 mag. 
These superoutburst intervals implies the supercycle of two years.
If this inference is correct, we can calculate the averaged supercycle of 728(7) d.
\citet{lev15amcvn} empirically obtained the following relation between the outburst recurrence time $\Delta T$ and $P_{\rm orb}$:
\begin{equation}
\Delta T {\rm[day]} = 1.53 \times 10^{-9} P_{\rm orb}{\rm [min]}^{7.35} + 24.7.
\end{equation}
This equation suggests that the supercycle of NSV 1440 is 474.6 d.
This value is roughly consistent with our estimation.
However, there is some uncertainty.
Although ASAS-3 had observed around NSV 1440 for nine years, the detected outburst is only two.
It might be caused by the shallow limiting magnitude of $V \sim 14$ \citep{ASAS3-2} and/or some observation gaps.
We should correct the supercycle by further observations.
}

\section{Summary}
The outbursts of NSV 1440 showed the following features, are known in H-rich WZ Sge-type DNe:
(1) double superoutbursts;
(2) early superhumps;
(3) late-stage superhumps instead of stage C ones;
(4) orbital-period modulations in the fading stage;
(5) multiple rebrightenings.
Therefore, we interpreted NSV 1440 as the first WZ Sge-type DN in AM CVn \tc{stars and candidates}.
This discovery implies that many AM CVn stars also show WZ Sge-type superoutbursts.
We should note that early and ordinary superhumps in WZ Sge-type superoutburst tell us
the basic binary parameters (i.e., orbital period and mass ratio).

We obtained the early superhump period of 0.0252329(49)
and the stage A superhump one of 0.025679(20) d from the 2015 outburst.
On the basis of these periods and the method of \citet{kat13qfromstageA}, we estimated $q=0.045(2)$.
\tc{This value suggests that the object is an AM CVn star and has a semi-degenerate secondary.
In other words, the object can be} a helium-star or evolved-CV channel system.
\tc{However, we should note that the validity of the $q$ estimation method from stage A $P_{\rm SH}$ is not confirmed in AM CVn stars.
Therefore, we need to compare $q$ from stage A with $q$ from other methods also in AM CVn stars.}

\section*{Supplementary Material}
The following supplementary data is available in the online article. figure E1--E2 and tables E1--E4.

\section*{Acknowledgments}
This work was supported by the Grant-in-Aid for Japan Society
for the Promotion of Science (JSPS) Fellows (No. 17J10039).
This work was also partially supported by the Grant-in-Aid “Initiative for
High-Dimensional Data-Driven Science through Deepening of Sparse Modeling” (25120007)
from the Ministry of Education, Culture, Sports, Science and Technology (MEXT) of Japan.
We are also thankful to the AAVSO International Database contributed by many worldwide observers
and to the survey project ASAS-SN.
This work has made use of data from the European Space Agency (ESA) mission
{\it Gaia} (https://www.cosmos.esa.int/gaia), processed by the {\it Gaia}
Data Processing and Analysis Consortium (DPAC,
https://www.cosmos.esa.int/web/gaia/dpac/consortium). Funding for the DPAC
has been provided by national institutions, in particular the institutions
participating in the {\it Gaia} Multilateral Agreement.

\bibliography{pasjadd,cvs,add}
\bibliographystyle{pasjtest1}

\setcounter{figure}{0}
\setcounter{table}{0}

\renewcommand{\thetable}{E\arabic{table}}
\renewcommand{\thefigure}{E\arabic{figure}}

\clearpage

\begin{figure*}[htb]
\begin{center}
    \FigureFile(160mm,100mm){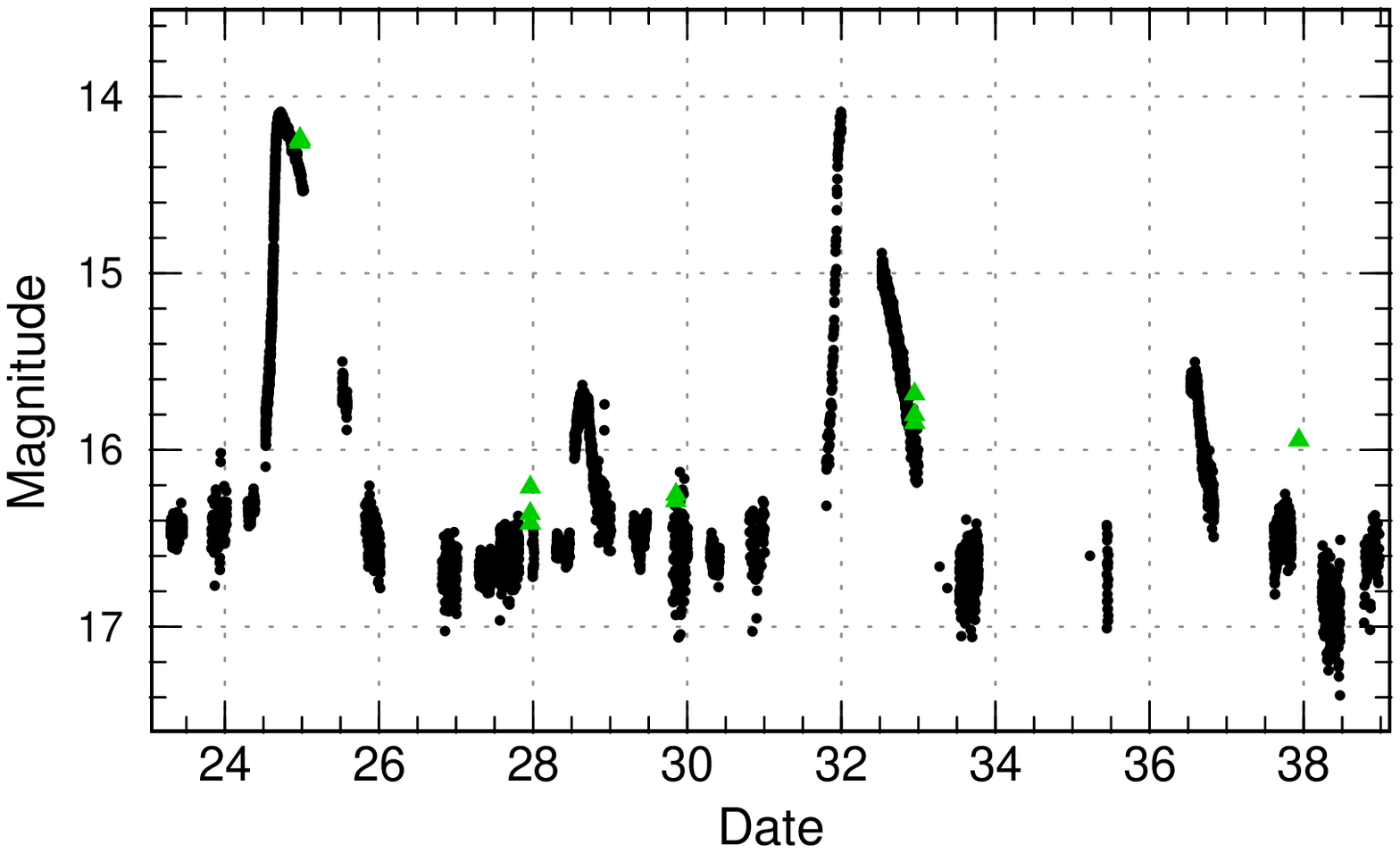}
\end{center}
\vspace{5mm}
\caption{Enlarged light curve of the 2015 outburst.
The object showed multiple rebrightenings and ``short rebrightenings''.
Circles and triangles represent our observations and the ASAS-SN $V$-band data, respectively.
The date is defined to be BJD $-$ 2457346.751.
} 
\label{fig:reb}
\end{figure*}

\begin{figure*}[htb]
\begin{center}
    \FigureFile(140mm,100mm){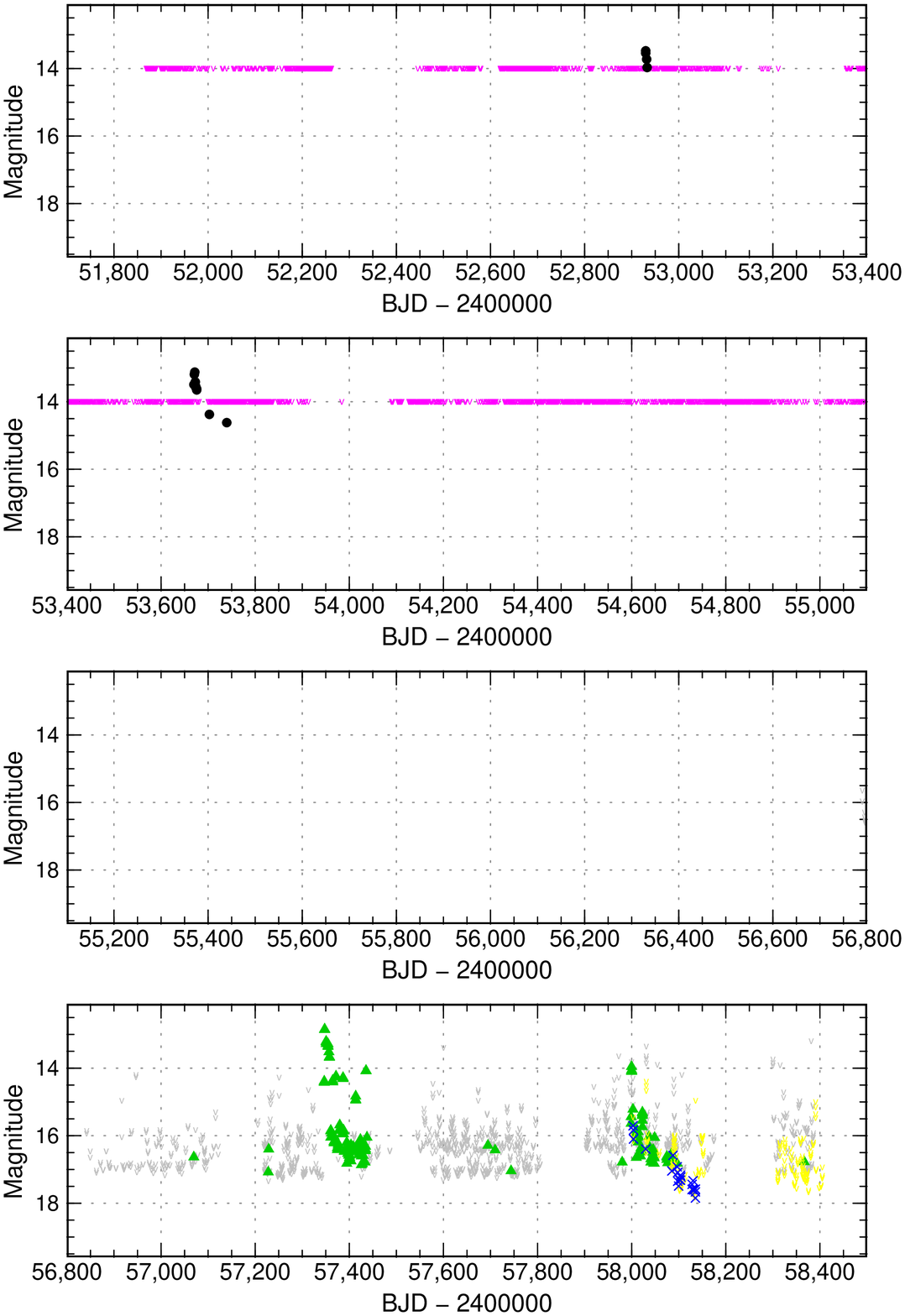}
\end{center}
\vspace{5mm}
\caption{The public light curves of ASAS-3 (Pojma\'nski 2002) and ASAS-SN (Shappee et al. 2014; Kochanek et al. 2017).
The black circles and purple ``V''-shapes represent the ASAS-3 data and its rough upper limits of $V=14$, respectively.
The green triangles and gray ``V''-shapes represent ASAS-SN $V$-band data and its upper limits, respectively.
The blue crosses and yellow ``V''-shapes represent ASAS-SN $g$-band data and its upper limits, respectively.
The object has shown four major outbursts on BJD 2452929.74 (in 2003), 2453669.80 (in 2005),  BJD 2457339.62 (in 2015) and BJD 2457987.27 (in 2017).
}  
\label{fig:reb}
\end{figure*}

\clearpage

\setcounter{table}{0}
\begin{table*}
\caption{Log of observations of NSV 1440 in 2015}
\label{tab:obs}
\begin{center}
\begin{tabular}{cccccc}
\hline
Start \commenta & End \commenta & Mag\commentb & $\sigma_{\rm Mag}$\commentc & $N$\commentd & Obs\commente \\ 
\hline
0.8898 & 1.0540 & 12.998 & 0.002 & 65 & HaC  \\
1.8870 & 2.0904 & 13.137 & 0.002 & 85 & HaC  \\
2.5267 & 2.7838 & 0.372 & 0.001 & 733 & MLF  \\
2.8842 & 3.0924 & 13.248 & 0.003 & 87 & HaC  \\
3.3320 & 3.4302 & 13.398 & 0.001 & 100 & SPET  \\
3.5174 & 3.6244 & 0.500 & 0.001 & 309 & MLF  \\
3.8813 & 4.0827 & 13.376 & 0.002 & 84 & HaC  \\
4.2724 & 4.3770 & 13.487 & 0.001 & 100 & SPET  \\
4.5354 & 4.7701 & 0.674 & 0.002 & 676 & MLF  \\
4.8785 & 5.0849 & 14.267 & 0.019 & 86 & HaC  \\
5.2262 & 5.3156 & 14.804 & 0.010 & 100 & COO  \\
5.2483 & 5.4314 & 15.262 & 0.012 & 154 & SPET  \\
5.8758 & 6.0843 & 17.085 & 0.048 & 86 & HaC  \\
6.8731 & 7.0832 & 16.084 & 0.022 & 70 & HaC  \\
7.5648 & 7.8429 & 0.836 & 0.005 & 802 & MLF  \\
7.8704 & 8.0834 & 13.465 & 0.004 & 115 & HaC  \\
8.5238 & 8.7154 & 0.604 & 0.002 & 552 & MLF  \\
8.8675 & 9.0839 & 13.481 & 0.004 & 118 & HaC  \\
10.0357 & 10.0840 & 13.665 & 0.005 & 29 & HaC  \\
10.5446 & 10.8418 & 0.916 & 0.001 & 855 & MLF  \\
11.8599 & 12.0832 & 13.959 & 0.003 & 120 & HaC  \\
12.2777 & 12.2777 & 14.153 & -- & 1 & SPET  \\
12.8571 & 13.0825 & 15.720 & 0.012 & 123 & HaC  \\
13.2295 & 13.3184 & 15.920 & 0.008 & 98 & COO  \\
13.5190 & 13.6784 & 3.160 & 0.005 & 401 & MLF  \\
13.8544 & 14.0833 & 16.084 & 0.008 & 125 & HaC  \\
14.1800 & 14.4771 & 16.225 & 0.003 & 315 & MGW  \\
14.2019 & 14.2837 & 16.082 & 0.006 & 93 & COO  \\
14.5276 & 14.7287 & 3.369 & 0.003 & 576 & MLF  \\
14.8515 & 15.0831 & 16.340 & 0.006 & 174 & HaC  \\
15.1737 & 15.4781 & 16.321 & 0.003 & 322 & MGW  \\
15.5221 & 15.6718 & 3.353 & 0.005 & 361 & MLF  \\
15.8489 & 16.0828 & 16.209 & 0.004 & 156 & HaC  \\
16.1631 & 16.4782 & 16.200 & 0.003 & 346 & MGW  \\
16.6098 & 16.8054 & 3.266 & 0.003 & 561 & MLF  \\
16.8461 & 17.0470 & 16.063 & 0.005 & 133 & HaC  \\
17.7688 & 17.8459 & 1.422 & 0.003 & 222 & MLF  \\
\hline
  \multicolumn{6}{l}{\commenta BJD$-$2457346.751 (same as figure 1, 3 and E1} \\
  \multicolumn{6}{l}{\commentb Mean magnitude. All observations are no filter (clear).} \\
  \multicolumn{6}{l}{\commentb Standard deviation of the observed magnitude.} \\
  \multicolumn{6}{l}{\commentd Number of observations.} \\
  \multicolumn{6}{l}{\commente Observer's code: HaC(F.-J. Hambsch), MLF (B. Monard),}\\
  \multicolumn{6}{l}{ MGW(G. Myers), SPET(P. Starr), COO(L. M. Cook)}\\
\end{tabular}
\end{center}
\end{table*}

\setcounter{table}{0}
\begin{table*}
\caption{Log of observations of NSV 1440 in 2015 (continued)}
\label{tab:obs}
\begin{center}
\begin{tabular}{cccccc}
\hline
Start \commenta & End \commenta & Mag\commentb & $\sigma_{\rm Mag}$\commentc & $N$\commentd & Obs\commente \\
\hline
17.8433 & 18.0439 & 14.546 & 0.011 & 122 & HaC  \\
18.8405 & 19.0404 & 16.448 & 0.009 & 124 & HaC  \\
19.8378 & 20.0388 & 16.529 & 0.007 & 125 & HaC  \\
20.2918 & 20.4782 & 16.484 & 0.003 & 214 & MGW  \\
20.5318 & 20.8392 & 3.514 & 0.002 & 878 & MLF  \\
20.8350 & 21.0360 & 16.455 & 0.006 & 125 & HaC  \\
21.8322 & 22.0319 & 16.607 & 0.007 & 120 & HaC  \\
22.2789 & 22.4782 & 16.564 & 0.004 & 244 & MGW  \\
22.8294 & 22.8473 & 16.599 & 0.026 & 9 & HaC  \\
23.2991 & 23.4568 & 16.440 & 0.003 & 164 & MGW  \\
23.8266 & 24.0277 & 16.414 & 0.009 & 123 & HaC  \\
24.2995 & 24.3904 & 16.324 & 0.004 & 115 & MGW  \\
24.5284 & 24.8259 & 2.061 & 0.025 & 827 & MLF  \\
24.8239 & 25.0247 & 14.368 & 0.009 & 123 & HaC  \\
25.5257 & 25.5913 & 2.859 & 0.009 & 58 & MLF  \\
25.8211 & 26.0214 & 16.509 & 0.010 & 123 & HaC  \\
26.8183 & 27.0186 & 16.695 & 0.010 & 119 & HaC  \\
27.2992 & 27.4784 & 16.670 & 0.004 & 213 & MGW  \\
27.5710 & 27.8199 & 3.752 & 0.005 & 353 & MLF  \\
27.9891 & 28.0160 & 16.590 & 0.020 & 19 & HaC  \\
28.3005 & 28.4782 & 16.547 & 0.005 & 106 & MGW  \\
28.5371 & 28.8419 & 3.069 & 0.009 & 431 & MLF  \\
28.8133 & 29.0129 & 16.356 & 0.010 & 121 & HaC  \\
29.3037 & 29.4775 & 16.473 & 0.005 & 175 & MGW  \\
29.8105 & 30.0106 & 16.594 & 0.016 & 121 & HaC  \\
30.2989 & 30.4215 & 16.606 & 0.005 & 119 & MGW  \\
30.8084 & 31.0071 & 16.548 & 0.017 & 74 & HaC  \\
31.8051 & 32.0047 & 15.364 & 0.067 & 86 & HaC  \\
32.5268 & 32.8433 & 2.488 & 0.007 & 907 & MLF  \\
32.8383 & 33.0023 & 15.943 & 0.016 & 75 & HaC  \\
33.2758 & 33.3766 & 16.825 & 0.082 & 2 & SPET  \\
33.5329 & 33.7819 & 3.851 & 0.006 & 352 & MLF  \\
35.2284 & 35.2284 & 16.701 & -- & 1 & SPET  \\
35.4451 & 35.4665 & 16.591 & 0.036 & 25 & COO  \\
36.5335 & 36.8398 & 3.123 & 0.012 & 433 & MLF  \\
37.6134 & 37.8440 & 3.639 & 0.005 & 326 & MLF  \\
38.2401 & 38.4775 & 16.884 & 0.008 & 280 & MGW  \\
38.2818 & 38.2818 & 16.755 & -- & 1 & SPET  \\
38.7856 & 38.9860 & 16.601 & 0.016 & 76 & HaC  \\
39.3522 & 39.3522 & 16.674 & -- & 1 & SPET  \\
39.4124 & 39.4778 & 16.637 & 0.007 & 81 & MGW  \\
39.7829 & 39.9810 & 15.389 & 0.080 & 63 & HaC  \\
40.2406 & 40.4519 & 14.697 & 0.010 & 225 & MGW  \\
\hline
\end{tabular}
\end{center}
\end{table*}

\setcounter{table}{0}
\begin{table*}
\caption{Log of observations of NSV 1440 in 2015 (continued)}
\label{tab:obs}
\begin{center}
\begin{tabular}{cccccc}
\hline
Start \commenta & End \commenta & Mag\commentb & $\sigma_{\rm Mag}$\commentc & $N$\commentd & Obs\commente \\
\hline
40.4201 & 40.4201 & 14.942 & 0.003 & 2 & SPET  \\
40.7802 & 40.9804 & 15.923 & 0.018 & 62 & HaC  \\
41.2516 & 41.4007 & 16.733 & 0.005 & 161 & MGW  \\
41.7775 & 41.9782 & 16.769 & 0.019 & 55 & HaC  \\
42.7766 & 42.9778 & 16.769 & 0.017 & 55 & HaC  \\
44.7712 & 44.9718 & 16.754 & 0.014 & 55 & HaC  \\
45.7628 & 45.9666 & 16.805 & 0.011 & 79 & HaC  \\
46.7644 & 46.9652 & 16.800 & 0.010 & 77 & HaC  \\
47.7702 & 47.9638 & 16.640 & 0.013 & 53 & HaC  \\
48.2146 & 48.4468 & 15.436 & 0.003 & 282 & MGW  \\
49.7638 & 49.9580 & 16.770 & 0.018 & 56 & HaC  \\
50.2256 & 50.4782 & 16.856 & 0.004 & 193 & MGW  \\
53.2775 & 53.4783 & 16.896 & 0.003 & 225 & MGW  \\
57.1948 & 57.3692 & 16.562 & 0.004 & 219 & MGW  \\
58.1977 & 58.4472 & 15.420 & 0.010 & 287 & MGW  \\
64.1787 & 64.3493 & 16.731 & 0.005 & 210 & MGW  \\
\hline
\end{tabular}
\end{center}
\end{table*}

\setcounter{table}{1}
\begin{table*}
\caption{Log of observations of NSV 1440 in 2017}
\label{tab:obs}
\begin{center}
\begin{tabular}{cccccc}
\hline
Start \commenta & End \commenta & Mag\commentb & $\sigma_{\rm Mag}$\commentc & $N$\commentd & Obs\commente \\ 
\hline
5.7946 & 5.8998 & 14.274 & 0.009 & 44 & HaC  \\
6.7918 & 6.8995 & 15.416 & 0.011 & 45 & HaC  \\
7.7885 & 7.8979 & 13.302 & 0.003 & 82 & HaC  \\
8.7857 & 8.8978 & 13.393 & 0.003 & 84 & HaC  \\
9.5052 & 9.6732 & 0.624 & 0.001 & 484 & MLF  \\
9.7823 & 9.8979 & 13.538 & 0.003 & 103 & HaC  \\
10.7800 & 10.8965 & 13.675 & 0.003 & 95 & HaC  \\
11.3870 & 11.6694 & 0.918 & 0.001 & 807 & MLF  \\
11.7772 & 11.8962 & 13.828 & 0.002 & 97 & HaC  \\
12.7744 & 12.9023 & 14.108 & 0.007 & 105 & HaC  \\
13.7716 & 13.9018 & 15.973 & 0.010 & 105 & HaC  \\
14.7689 & 14.9003 & 16.184 & 0.010 & 98 & HaC  \\
15.7663 & 15.9000 & 16.249 & 0.008 & 86 & HaC  \\
16.7636 & 16.8988 & 16.222 & 0.007 & 87 & HaC  \\
17.7608 & 17.8988 & 14.229 & 0.005 & 89 & HaC  \\
19.7553 & 19.8975 & 16.566 & 0.013 & 92 & HaC  \\
20.7246 & 20.8958 & 16.400 & 0.022 & 76 & HaC  \\
21.7218 & 21.8959 & 16.457 & 0.012 & 77 & HaC  \\
25.3862 & 25.6586 & 2.588 & 0.007 & 758 & MLF  \\
25.7684 & 25.8820 & 16.065 & 0.012 & 64 & HaC  \\
26.7656 & 26.8219 & 16.670 & 0.033 & 28 & HaC  \\
28.7600 & 28.8910 & 16.515 & 0.009 & 77 & HaC  \\
29.7573 & 29.8910 & 16.488 & 0.008 & 78 & HaC  \\
32.9011 & 32.9018 & 15.777 & 0.176 & 2 & HaC  \\
33.9004 & 33.9011 & 16.713 & 0.121 & 2 & HaC  \\
34.8998 & 34.9006 & 16.710 & 0.049 & 2 & HaC  \\
35.8990 & 35.8997 & 16.585 & 0.047 & 2 & HaC  \\
36.8983 & 36.8991 & 16.733 & 0.199 & 2 & HaC  \\
37.8976 & 37.8976 & 16.551 & -- & 1 & HaC  \\
38.8970 & 38.8970 & 16.670 & -- & 1 & HaC  \\
39.8944 & 39.8952 & 14.891 & 0.006 & 2 & HaC  \\
40.8903 & 40.8910 & 15.913 & 0.046 & 2 & HaC  \\
41.3925 & 41.5975 & 3.957 & 0.005 & 510 & MLF  \\
41.8896 & 41.8903 & 16.777 & 0.171 & 2 & HaC  \\
42.8889 & 42.8897 & 16.868 & 0.182 & 2 & HaC  \\
43.8882 & 43.8889 & 16.759 & 0.015 & 2 & HaC  \\
44.8875 & 44.8882 & 16.665 & 0.011 & 2 & HaC  \\
\hline
  \multicolumn{6}{l}{\commenta BJD$-$2457982.000 (same as figure 1 and 3} \\
  \multicolumn{6}{l}{\commentb Mean magnitude. All observations are no filter (clear).} \\
  \multicolumn{6}{l}{\commentb Standard deviation of the observed magnitude.} \\
  \multicolumn{6}{l}{\commentd Number of observations.} \\
  \multicolumn{6}{l}{\commente Observer's code: HaC(F.-J. Hambsch), MLF (B. Monard),}\\
  \multicolumn{6}{l}{ MGW(G. Myers), SPET(P. Starr), COO(L. M. Cook)}\\
\end{tabular}
\end{center}
\end{table*}

\setcounter{table}{1}
\begin{table*}
\caption{Log of observations of NSV 1440 in 2017 (continued)}
\label{tab:obs}
\begin{center}
\begin{tabular}{cccccc}
\hline
Start \commenta & End \commenta & Mag\commentb & $\sigma_{\rm Mag}$\commentc & $N$\commentd & Obs\commente \\
\hline
45.8868 & 45.8875 & 16.726 & 0.051 & 2 & HaC  \\
46.8861 & 46.8869 & 16.815 & 0.023 & 2 & HaC  \\
47.8854 & 47.8862 & 16.697 & 0.091 & 2 & HaC  \\
48.8847 & 48.8855 & 15.685 & 0.068 & 2 & HaC  \\
49.8840 & 49.8848 & 15.976 & 0.090 & 2 & HaC  \\
50.8833 & 50.8841 & 16.489 & 0.012 & 2 & HaC  \\
51.8826 & 51.8834 & 16.604 & 0.210 & 2 & HaC  \\
52.8820 & 52.8827 & 16.841 & 0.125 & 2 & HaC  \\
53.8812 & 53.8820 & 16.546 & 0.009 & 2 & HaC  \\
54.8805 & 54.8813 & 16.529 & 0.086 & 2 & HaC  \\
55.8799 & 55.8806 & 16.531 & 0.080 & 2 & HaC  \\
56.8791 & 56.8799 & 15.122 & 0.008 & 2 & HaC  \\
57.8793 & 57.8793 & 17.926 & -- & 1 & HaC  \\
58.8777 & 58.8785 & 16.718 & 0.190 & 2 & HaC  \\
59.8771 & 59.8778 & 16.812 & 0.014 & 2 & HaC  \\
60.8763 & 60.8771 & 16.955 & 0.163 & 2 & HaC  \\
61.8757 & 61.8764 & 16.825 & 0.103 & 2 & HaC  \\
62.8750 & 62.8757 & 16.812 & 0.166 & 2 & HaC  \\
63.8743 & 63.8751 & 16.824 & 0.142 & 2 & HaC  \\
64.8736 & 64.8743 & 16.719 & 0.020 & 2 & HaC  \\
65.8728 & 65.8736 & 16.655 & 0.044 & 2 & HaC  \\
66.8722 & 66.8730 & 16.689 & 0.090 & 2 & HaC  \\
67.8715 & 67.8722 & 14.906 & 0.018 & 2 & HaC  \\
68.8715 & 68.8722 & 16.942 & 0.140 & 2 & HaC  \\
69.8708 & 69.8715 & 16.970 & 0.104 & 2 & HaC  \\
70.8702 & 70.8709 & 16.777 & 0.082 & 2 & HaC  \\
71.8694 & 71.8701 & 17.052 & 0.250 & 2 & HaC  \\
72.8688 & 72.8695 & 16.920 & 0.139 & 2 & HaC  \\
73.8680 & 73.8687 & 16.940 & 0.040 & 2 & HaC  \\
74.8681 & 74.8688 & 17.016 & 0.088 & 2 & HaC  \\
75.8674 & 75.8681 & 16.809 & 0.164 & 2 & HaC  \\
76.8667 & 76.8674 & 16.901 & 0.113 & 2 & HaC  \\
77.8659 & 77.8667 & 16.988 & 0.088 & 2 & HaC  \\
78.8660 & 78.8667 & 17.099 & 0.495 & 2 & HaC  \\
79.8653 & 79.8660 & 15.047 & 0.037 & 2 & HaC  \\
80.8646 & 80.8653 & 16.910 & 0.030 & 2 & HaC  \\
81.8639 & 81.8646 & 16.958 & 0.251 & 2 & HaC  \\
82.6613 & 82.6621 & 16.957 & 0.113 & 2 & HaC  \\
83.6586 & 83.6593 & 16.945 & 0.020 & 2 & HaC  \\
84.6558 & 84.6566 & 17.101 & 0.173 & 2 & HaC  \\
85.3592 & 85.4434 & 4.394 & 0.011 & 193 & MLF  \\
85.6530 & 85.6537 & 17.335 & 0.180 & 2 & HaC  \\
86.6502 & 86.6509 & 17.080 & 0.033 & 2 & HaC  \\
\hline
\end{tabular}
\end{center}
\end{table*}

\setcounter{table}{1}
\begin{table*}
\caption{Log of observations of NSV 1440 in 2017 (continued)}
\label{tab:obs}
\begin{center}
\begin{tabular}{cccccc}
\hline
Start \commenta & End \commenta & Mag\commentb & $\sigma_{\rm Mag}$\commentc & $N$\commentd & Obs\commente \\
\hline
87.6475 & 87.6482 & 17.247 & 0.049 & 2 & HaC  \\
88.6447 & 88.6454 & 17.172 & 0.326 & 2 & HaC  \\
89.6419 & 89.6426 & 17.245 & 0.222 & 2 & HaC  \\
90.6391 & 90.6398 & 17.017 & 0.034 & 2 & HaC  \\
91.6363 & 91.6371 & 17.209 & 0.118 & 2 & HaC  \\
92.6336 & 92.6343 & 17.052 & 0.060 & 2 & HaC  \\
93.6307 & 93.6314 & 17.212 & 0.247 & 2 & HaC  \\
94.6279 & 94.6287 & 17.276 & 0.096 & 2 & HaC  \\
95.6252 & 95.6259 & 17.181 & 0.044 & 2 & HaC  \\
96.6224 & 96.6231 & 17.173 & 0.156 & 2 & HaC  \\
97.6197 & 97.6204 & 17.237 & 0.164 & 2 & HaC  \\
98.6169 & 98.6176 & 17.326 & 0.055 & 2 & HaC  \\
99.6142 & 99.6149 & 17.835 & 0.132 & 2 & HaC  \\
100.6120 & 100.6128 & 17.216 & 0.039 & 2 & HaC  \\
101.6093 & 101.6100 & 17.221 & 0.070 & 2 & HaC  \\
102.6065 & 102.6072 & 17.676 & 0.175 & 2 & HaC  \\
103.6037 & 103.6044 & 17.229 & 0.058 & 2 & HaC  \\
104.6008 & 104.6016 & 17.435 & 0.461 & 2 & HaC  \\
105.5981 & 105.5988 & 17.258 & 0.250 & 2 & HaC  \\
106.5953 & 106.5960 & 17.167 & 0.210 & 2 & HaC  \\
107.5926 & 107.5933 & 17.446 & 0.021 & 2 & HaC  \\
108.5899 & 108.5906 & 17.193 & 0.002 & 2 & HaC  \\
109.5870 & 109.5877 & 16.945 & 0.241 & 2 & HaC  \\
111.5815 & 111.5822 & 17.725 & 0.234 & 2 & HaC  \\
113.5759 & 113.5767 & 17.968 & 0.194 & 2 & HaC  \\
114.5732 & 114.5739 & 18.137 & 0.137 & 2 & HaC  \\
115.5704 & 115.5711 & 17.611 & 0.060 & 2 & HaC  \\
116.5676 & 116.5684 & 18.074 & 0.186 & 2 & HaC  \\
117.5648 & 117.5655 & 17.839 & 0.136 & 2 & HaC  \\
118.5620 & 118.5628 & 18.186 & 0.224 & 2 & HaC  \\
119.5592 & 119.5600 & 18.203 & 0.189 & 2 & HaC  \\
120.5565 & 120.5572 & 18.481 & 0.097 & 2 & HaC  \\
121.5538 & 121.5545 & 18.453 & 0.254 & 2 & HaC  \\
122.5516 & 122.5524 & 17.954 & 0.265 & 2 & HaC  \\
125.5578 & 125.5586 & 18.803 & 0.426 & 2 & HaC  \\
126.5558 & 126.5565 & 18.530 & 0.096 & 2 & HaC  \\
127.5537 & 127.5537 & 18.328 & -- & 1 & HaC  \\
128.5503 & 128.5510 & 18.204 & 0.139 & 2 & HaC  \\
131.5410 & 131.5410 & 18.045 & -- & 1 & HaC  \\
133.5347 & 133.5347 & 18.407 & -- & 1 & HaC  \\
137.5244 & 137.5244 & 17.673 & -- & 1 & HaC  \\
142.5126 & 142.5126 & 18.037 & -- & 1 & HaC  \\
\hline
\end{tabular}
\end{center}
\end{table*}

\setcounter{table}{2}
\begin{table*}
\caption{Timings of superhump maxima of NSV 1440 in 2015.}
\begin{center}
\begin{tabular}{ccccc}
\hline
$E$ & Maximum time\commenta & Error & $O-C$\commentb & $N$\commentc\\
\hline
0 & 7.03553 & 0.00068 & -0.00798 & 7 \\
1 & 7.05998 & 0.00058 & -0.00890 & 7 \\
21 & 7.57272 & 0.00061 & -0.00330 & 47 \\
24 & 7.65045 & 0.00051 & -0.00164 & 60 \\
25 & 7.67702 & 0.00050 & -0.00044 & 58 \\
26 & 7.70118 & 0.00061 & -0.00162 & 58 \\
27 & 7.72639 & 0.00042 & -0.00178 & 58 \\
28 & 7.75210 & 0.00053 & -0.00143 & 59 \\
29 & 7.78111 & 0.00097 & 0.00223 & 58 \\
30 & 7.80405 & 0.00046 & -0.00019 & 58 \\
31 & 7.82826 & 0.00049 & -0.00134 & 59 \\
38 & 8.00756 & 0.00106 & 0.00046 & 12 \\
59 & 8.54141 & 0.00019 & 0.00180 & 58 \\
60 & 8.56712 & 0.00016 & 0.00215 & 58 \\
61 & 8.59260 & 0.00018 & 0.00227 & 58 \\
62 & 8.61800 & 0.00022 & 0.00232 & 59 \\
63 & 8.64317 & 0.00019 & 0.00214 & 59 \\
64 & 8.66806 & 0.00021 & 0.00166 & 58 \\
65 & 8.69376 & 0.00019 & 0.00201 & 58 \\
73 & 8.89717 & 0.00082 & 0.00255 & 12 \\
76 & 8.97323 & 0.00037 & 0.00254 & 11 \\
79 & 9.04841 & 0.00064 & 0.00165 & 11 \\
139 & 10.57123 & 0.00039 & 0.00302 & 59 \\
140 & 10.59640 & 0.00034 & 0.00283 & 59 \\
141 & 10.62099 & 0.00035 & 0.00207 & 58 \\
142 & 10.64665 & 0.00024 & 0.00237 & 59 \\
143 & 10.67219 & 0.00029 & 0.00255 & 59 \\
144 & 10.69760 & 0.00021 & 0.00260 & 58 \\
145 & 10.72232 & 0.00047 & 0.00196 & 59 \\
146 & 10.74784 & 0.00035 & 0.00212 & 59 \\
147 & 10.77259 & 0.00035 & 0.00152 & 59 \\
148 & 10.79878 & 0.00035 & 0.00235 & 58 \\
149 & 10.82362 & 0.00035 & 0.00183 & 59 \\
191 & 11.88719 & 0.00078 & 0.00038 & 12 \\
192 & 11.91419 & 0.00091 & 0.00202 & 10 \\
193 & 11.93871 & 0.00061 & 0.00119 & 10 \\
194 & 11.96480 & 0.00038 & 0.00193 & 12 \\
195 & 11.98758 & 0.00089 & -0.00066 & 11 \\
196 & 12.01453 & 0.00089 & 0.00093 & 12 \\
197 & 12.04180 & 0.00066 & 0.00285 & 11 \\
230 & 12.88009 & 0.00088 & 0.00434 & 9 \\
\hline
  \multicolumn{5}{l}{\commenta BJD$-$2457346.751 (same as figure 1 and 3).} \\
  \multicolumn{5}{l}{\commentb  $C= 2457353.794513 + 0.02535754750 E$.} \\
  \multicolumn{5}{l}{\commentc Number of points used to determine the maximum.} \\
\end{tabular}
\end{center}
\end{table*}

\setcounter{table}{2}
\begin{table*}
\caption{Timings of superhump maxima of NSV 1440 in 2015 (continued).}
\begin{center}

\begin{tabular}{ccccc}
\hline
$E$ & Maximum time\commenta & Error & $O-C$\commentb & $N$\commentc\\
\hline
232 & 12.92823 & 0.00084 & 0.00176 & 10 \\
233 & 12.95554 & 0.00054 & 0.00371 & 12 \\
234 & 12.98005 & 0.00064 & 0.00287 & 12 \\
235 & 13.00527 & 0.00053 & 0.00274 & 11 \\
236 & 13.03049 & 0.00094 & 0.00260 & 11 \\
237 & 13.05637 & 0.00049 & 0.00312 & 12 \\
244 & 13.23305 & 0.00053 & 0.00230 & 20 \\
245 & 13.25899 & 0.00040 & 0.00288 & 23 \\
246 & 13.28529 & 0.00100 & 0.00382 & 21 \\
247 & 13.30968 & 0.00105 & 0.00285 & 18 \\
256 & 13.53975 & 0.00047 & 0.00470 & 54 \\
257 & 13.56191 & 0.00041 & 0.00151 & 49 \\
259 & 13.60964 & 0.00101 & -0.00148 & 59 \\
261 & 13.66484 & 0.00060 & 0.00301 & 54 \\
283 & 14.21677 & 0.00040 & -0.00293 & 47 \\
284 & 14.24298 & 0.00040 & -0.00208 & 48 \\
287 & 14.31922 & 0.00309 & -0.00191 & 25 \\
288 & 14.34528 & 0.00112 & -0.00121 & 24 \\
289 & 14.37048 & 0.00112 & -0.00137 & 24 \\
290 & 14.39580 & 0.00122 & -0.00140 & 24 \\
292 & 14.44656 & 0.00078 & -0.00136 & 24 \\
293 & 14.47320 & 0.00098 & -0.00007 & 19 \\
296 & 14.54730 & 0.00130 & -0.00205 & 59 \\
297 & 14.57497 & 0.00094 & 0.00026 & 56 \\
298 & 14.59405 & 0.00087 & -0.00601 & 59 \\
299 & 14.61735 & 0.00086 & -0.00807 & 58 \\
300 & 14.64042 & 0.00641 & -0.01036 & 58 \\
301 & 14.67064 & 0.00067 & -0.00550 & 58 \\
302 & 14.70126 & 0.00134 & -0.00024 & 59 \\
322 & 15.20068 & 0.00060 & -0.00796 & 26 \\
323 & 15.22620 & 0.00109 & -0.00780 & 25 \\
327 & 15.32919 & 0.00099 & -0.00624 & 25 \\
328 & 15.35461 & 0.00168 & -0.00618 & 25 \\
329 & 15.37861 & 0.00232 & -0.00754 & 25 \\
331 & 15.43101 & 0.00057 & -0.00585 & 26 \\
332 & 15.45570 & 0.00080 & -0.00652 & 25 \\
336 & 15.55663 & 0.00082 & -0.00702 & 41 \\
337 & 15.58332 & 0.00096 & -0.00569 & 45 \\
338 & 15.60673 & 0.00106 & -0.00763 & 58 \\
339 & 15.63103 & 0.00064 & -0.00869 & 49 \\
340 & 15.65951 & 0.00096 & -0.00557 & 55 \\
349 & 15.88843 & 0.00106 & -0.00487 & 15 \\
353 & 15.98966 & 0.00082 & -0.00507 & 15 \\
354 & 16.01330 & 0.00078 & -0.00679 & 14 \\
\hline
\end{tabular}
\end{center}
\end{table*}

\setcounter{table}{2}
\begin{table*}
\caption{Timings of superhump maxima of NSV 1440 in 2015 (continued).}
\begin{center}
\begin{tabular}{ccccc}
\hline
$E$ & Maximum time\commenta & Error & $O-C$\commentb & $N$\commentc\\
\hline
355 & 16.03913 & 0.00089 & -0.00632 & 14 \\
356 & 16.06652 & 0.00088 & -0.00428 & 14 \\
361 & 16.19323 & 0.00044 & -0.00436 & 25 \\
362 & 16.21864 & 0.00048 & -0.00431 & 26 \\
363 & 16.24375 & 0.00094 & -0.00456 & 26 \\
366 & 16.31962 & 0.00046 & -0.00476 & 25 \\
367 & 16.34620 & 0.00076 & -0.00353 & 25 \\
368 & 16.37098 & 0.00040 & -0.00411 & 26 \\
369 & 16.39607 & 0.00052 & -0.00438 & 26 \\
370 & 16.42090 & 0.00071 & -0.00490 & 25 \\
371 & 16.44703 & 0.00035 & -0.00413 & 26 \\
378 & 16.62417 & 0.00077 & -0.00449 & 58 \\
379 & 16.64643 & 0.00125 & -0.00759 & 58 \\
380 & 16.67116 & 0.00078 & -0.00822 & 58 \\
381 & 16.70058 & 0.00114 & -0.00416 & 58 \\
382 & 16.72708 & 0.00097 & -0.00302 & 58 \\
383 & 16.75134 & 0.00095 & -0.00412 & 58 \\
384 & 16.77798 & 0.00051 & -0.00283 & 58 \\
389 & 16.90430 & 0.00076 & -0.00330 & 14 \\
392 & 16.98097 & 0.00169 & -0.00270 & 14 \\
394 & 17.02965 & 0.00103 & -0.00474 & 15 \\
523 & 20.30808 & 0.00082 & 0.00257 & 26 \\
524 & 20.33344 & 0.00104 & 0.00257 & 26 \\
525 & 20.35825 & 0.00087 & 0.00202 & 25 \\
526 & 20.38164 & 0.00092 & 0.00006 & 25 \\
527 & 20.40691 & 0.00044 & -0.00003 & 22 \\
529 & 20.45956 & 0.00119 & 0.00191 & 26 \\
546 & 20.89078 & 0.00057 & 0.00204 & 11 \\
547 & 20.91461 & 0.00091 & 0.00051 & 13 \\
601 & 22.29049 & 0.00048 & 0.00709 & 25 \\
602 & 22.31738 & 0.00080 & 0.00863 & 25 \\
603 & 22.34062 & 0.00081 & 0.00651 & 25 \\
604 & 22.36770 & 0.00087 & 0.00823 & 25 \\
605 & 22.39283 & 0.00070 & 0.00800 & 24 \\
606 & 22.41751 & 0.00112 & 0.00732 & 24 \\
607 & 22.44401 & 0.00062 & 0.00846 & 25 \\
608 & 22.46759 & 0.00054 & 0.00669 & 23 \\
641 & 23.30594 & 0.00051 & 0.00824 & 25 \\
642 & 23.33218 & 0.00040 & 0.00912 & 26 \\
643 & 23.35865 & 0.00042 & 0.01023 & 26 \\
662 & 23.83888 & 0.00099 & 0.00867 & 7 \\
664 & 23.88559 & 0.00052 & 0.00467 & 12 \\
681 & 24.32273 & 0.00052 & 0.01072 & 26 \\
682 & 24.34793 & 0.00048 & 0.01057 & 25 \\
\hline
\end{tabular}
\end{center}
\end{table*}

\setcounter{table}{2}
\begin{table*}
\caption{Timings of superhump maxima of NSV 1440 in 2015 (continued).}
\begin{center}

\begin{tabular}{ccccc}
\hline
$E$ & Maximum time\commenta & Error & $O-C$\commentb & $N$\commentc\\
\hline
683 & 24.37235 & 0.00059 & 0.00963 & 25 \\
799 & 27.32055 & 0.00052 & 0.01636 & 24 \\
800 & 27.34656 & 0.00066 & 0.01701 & 24 \\
801 & 27.37153 & 0.00043 & 0.01662 & 24 \\
802 & 27.39701 & 0.00036 & 0.01674 & 24 \\
803 & 27.42265 & 0.00041 & 0.01703 & 24 \\
804 & 27.44837 & 0.00063 & 0.01738 & 23 \\
805 & 27.47338 & 0.00069 & 0.01704 & 17 \\
810 & 27.60095 & 0.00064 & 0.01782 & 29 \\
811 & 27.62675 & 0.00085 & 0.01827 & 28 \\
812 & 27.65198 & 0.00087 & 0.01814 & 29 \\
813 & 27.67551 & 0.00065 & 0.01631 & 29 \\
814 & 27.70230 & 0.00078 & 0.01774 & 29 \\
815 & 27.72874 & 0.00110 & 0.01882 & 29 \\
817 & 27.78457 & 0.00207 & 0.02394 & 29 \\
818 & 27.80378 & 0.00056 & 0.01780 & 29 \\
838 & 28.31031 & 0.00074 & 0.01717 & 22 \\
843 & 28.43813 & 0.00058 & 0.01820 & 21 \\
844 & 28.46245 & 0.00066 & 0.01716 & 22 \\
855 & 28.73965 & 0.00085 & 0.01543 & 29 \\
856 & 28.76774 & 0.00065 & 0.01816 & 29 \\
857 & 28.79029 & 0.00120 & 0.01536 & 28 \\
858 & 28.82008 & 0.00267 & 0.01979 & 35 \\
878 & 29.32823 & 0.00090 & 0.02079 & 21 \\
879 & 29.35354 & 0.00075 & 0.02074 & 21 \\
880 & 29.37978 & 0.00061 & 0.02162 & 20 \\
881 & 29.40311 & 0.00028 & 0.01960 & 21 \\
882 & 29.42966 & 0.00076 & 0.02079 & 20 \\
883 & 29.45506 & 0.00086 & 0.02083 & 19 \\
898 & 29.83663 & 0.00365 & 0.02204 & 9 \\
900 & 29.88182 & 0.00122 & 0.01651 & 12 \\
902 & 29.93984 & 0.00141 & 0.02382 & 13 \\
904 & 29.98983 & 0.00081 & 0.02310 & 13 \\
917 & 30.31616 & 0.00072 & 0.01977 & 25 \\
918 & 30.34502 & 0.00052 & 0.02328 & 25 \\
919 & 30.36800 & 0.00051 & 0.02090 & 23 \\
1006 & 32.57241 & 0.00047 & 0.01920 & 59 \\
1007 & 32.59711 & 0.00076 & 0.01855 & 58 \\
1008 & 32.62250 & 0.00069 & 0.01858 & 58 \\
1009 & 32.65006 & 0.00115 & 0.02078 & 58 \\
1010 & 32.67470 & 0.00085 & 0.02006 & 58 \\
1011 & 32.70033 & 0.00052 & 0.02033 & 59 \\
1012 & 32.72316 & 0.00106 & 0.01781 & 59 \\
1013 & 32.74884 & 0.00058 & 0.01813 & 58 \\
\hline
\end{tabular}
\end{center}
\end{table*}

\setcounter{table}{2}
\begin{table*}
\caption{Timings of superhump maxima of NSV 1440 in 2015 (continued).}
\begin{center}

\begin{tabular}{ccccc}
\hline
$E$ & Maximum time\commenta & Error & $O-C$\commentb & $N$\commentc\\
\hline
1014 & 32.77472 & 0.00089 & 0.01865 & 58 \\
1015 & 32.80402 & 0.00089 & 0.02260 & 58 \\
1016 & 32.82728 & 0.00088 & 0.02050 & 58 \\
1162 & 36.54211 & 0.00108 & 0.03313 & 28 \\
1163 & 36.56420 & 0.00149 & 0.02985 & 28 \\
1164 & 36.59776 & 0.00303 & 0.03806 & 29 \\
1165 & 36.61692 & 0.00094 & 0.03186 & 28 \\
1166 & 36.63856 & 0.00146 & 0.02814 & 28 \\
1167 & 36.66798 & 0.00182 & 0.03220 & 27 \\
1168 & 36.68965 & 0.00060 & 0.02852 & 29 \\
1169 & 36.71854 & 0.00090 & 0.03206 & 29 \\
1170 & 36.74117 & 0.00090 & 0.02933 & 29 \\
1205 & 37.63432 & 0.00070 & 0.03496 & 29 \\
1206 & 37.65766 & 0.00085 & 0.03295 & 29 \\
1207 & 37.68302 & 0.00108 & 0.03295 & 29 \\
1208 & 37.70748 & 0.00031 & 0.03205 & 28 \\
1209 & 37.73136 & 0.00199 & 0.03057 & 29 \\
1210 & 37.75901 & 0.00093 & 0.03287 & 28 \\
1211 & 37.78580 & 0.00176 & 0.03429 & 29 \\
1212 & 37.81117 & 0.00129 & 0.03431 & 28 \\
1213 & 37.83164 & 0.00124 & 0.02942 & 27 \\
1236 & 38.41839 & 0.00139 & 0.03295 & 24 \\
1237 & 38.44155 & 0.00092 & 0.03076 & 25 \\
1238 & 38.46594 & 0.00271 & 0.02978 & 25 \\
1309 & 40.26518 & 0.00105 & 0.02864 & 23 \\
1311 & 40.31788 & 0.00129 & 0.03062 & 22 \\
1312 & 40.34172 & 0.00192 & 0.02911 & 23 \\
1315 & 40.42033 & 0.00103 & 0.03164 & 24 \\
1348 & 41.25810 & 0.00064 & 0.03261 & 20 \\
1349 & 41.28359 & 0.00076 & 0.03275 & 23 \\
1350 & 41.31109 & 0.00149 & 0.03489 & 23 \\
1351 & 41.33288 & 0.00085 & 0.03132 & 22 \\
1352 & 41.36051 & 0.00101 & 0.03360 & 23 \\
1353 & 41.38411 & 0.00060 & 0.03184 & 17 \\
1526 & 45.77633 & 0.00303 & 0.03720 & 13 \\
1625 & 48.29053 & 0.00140 & 0.04100 & 26 \\
1626 & 48.31561 & 0.00139 & 0.04072 & 27 \\
1627 & 48.34478 & 0.00124 & 0.04454 & 25 \\
1823 & 53.32038 & 0.00067 & 0.05006 & 25 \\
1824 & 53.34916 & 0.00064 & 0.05348 & 26 \\
1828 & 53.44638 & 0.00080 & 0.04927 & 25 \\
1978 & 57.24552 & 0.00083 & 0.04478 & 26 \\
1979 & 57.27226 & 0.00265 & 0.04616 & 26 \\
1980 & 57.29554 & 0.00070 & 0.04408 & 26 \\
\hline
\end{tabular}
\end{center}
\end{table*}

\setcounter{table}{2}
\begin{table*}
\caption{Timings of superhump maxima of NSV 1440 in 2015 (continued).}
\begin{center}

\begin{tabular}{ccccc}
\hline
$E$ & Maximum time\commenta & Error & $O-C$\commentb & $N$\commentc\\
\hline
1981 & 57.32261 & 0.00171 & 0.04580 & 26 \\
2016 & 58.21708 & 0.00049 & 0.05275 & 25 \\
2017 & 58.24373 & 0.00136 & 0.05404 & 26 \\
2018 & 58.26918 & 0.00102 & 0.05414 & 25 \\
2019 & 58.29248 & 0.00037 & 0.05207 & 26 \\
2020 & 58.31870 & 0.00034 & 0.05294 & 25 \\
2021 & 58.34415 & 0.00047 & 0.05303 & 26 \\
2022 & 58.36952 & 0.00042 & 0.05305 & 25 \\
2023 & 58.39516 & 0.00059 & 0.05333 & 22 \\
2251 & 64.18447 & 0.00064 & 0.06112 & 23 \\
2252 & 64.20992 & 0.00117 & 0.06121 & 24 \\
2255 & 64.28665 & 0.00140 & 0.06187 & 26 \\
2256 & 64.31203 & 0.00118 & 0.06189 & 25 \\
2257 & 64.33610 & 0.00119 & 0.06060 & 24 \\
\hline
\end{tabular}
\end{center}
\end{table*}

\setcounter{table}{3}
\begin{table*}
\caption{Timings of superhump maxima of NSV 1440 in 2017.}
\begin{center}

\begin{tabular}{ccccc}
\hline
$E$ & Maximum time\commenta & Error & $O-C$\commentb & $N$\commentc\\
\hline
30 & 7.79878 & 0.00067 & -0.00532 & 16 \\
31 & 7.82167 & 0.00071 & -0.00778 & 15 \\
32 & 7.84789 & 0.00070 & -0.00693 & 15 \\
33 & 7.87355 & 0.00048 & -0.00662 & 15 \\
34 & 7.89908 & 0.00098 & -0.00645 & 9 \\
70 & 8.81625 & 0.00050 & -0.00215 & 15 \\
71 & 8.84228 & 0.00070 & -0.00148 & 15 \\
72 & 8.86629 & 0.00098 & -0.00283 & 15 \\
73 & 8.89179 & 0.00044 & -0.00268 & 15 \\
98 & 9.52667 & 0.00029 & -0.00174 & 58 \\
99 & 9.55158 & 0.00026 & -0.00219 & 58 \\
100 & 9.57734 & 0.00032 & -0.00179 & 58 \\
101 & 9.60212 & 0.00025 & -0.00237 & 58 \\
102 & 9.62767 & 0.00023 & -0.00217 & 58 \\
103 & 9.65369 & 0.00027 & -0.00151 & 58 \\
109 & 9.80509 & 0.00118 & -0.00226 & 17 \\
110 & 9.83168 & 0.00108 & -0.00103 & 18 \\
111 & 9.85618 & 0.00080 & -0.00189 & 18 \\
112 & 9.88081 & 0.00048 & -0.00261 & 17 \\
149 & 10.82032 & 0.00119 & -0.00133 & 17 \\
150 & 10.84469 & 0.00049 & -0.00231 & 16 \\
151 & 10.87055 & 0.00044 & -0.00181 & 15 \\
172 & 11.40200 & 0.00034 & -0.00288 & 58 \\
173 & 11.42743 & 0.00038 & -0.00280 & 56 \\
174 & 11.45236 & 0.00034 & -0.00323 & 59 \\
175 & 11.47869 & 0.00048 & -0.00226 & 57 \\
176 & 11.50452 & 0.00029 & -0.00178 & 58 \\
177 & 11.52973 & 0.00052 & -0.00193 & 58 \\
178 & 11.55450 & 0.00048 & -0.00252 & 59 \\
179 & 11.57972 & 0.00034 & -0.00265 & 57 \\
180 & 11.60529 & 0.00038 & -0.00244 & 58 \\
181 & 11.63030 & 0.00030 & -0.00278 & 58 \\
182 & 11.65545 & 0.00047 & -0.00300 & 57 \\
187 & 11.78264 & 0.00052 & -0.00260 & 15 \\
188 & 11.80732 & 0.00054 & -0.00327 & 18 \\
189 & 11.83251 & 0.00052 & -0.00344 & 15 \\
190 & 11.86024 & 0.00121 & -0.00106 & 16 \\
191 & 11.88319 & 0.00112 & -0.00347 & 15 \\
227 & 12.79641 & 0.00056 & -0.00313 & 19 \\
228 & 12.82239 & 0.00036 & -0.00250 & 16 \\
\hline
  \multicolumn{5}{l}{\commenta BJD$-$2457982.000 (same as figure 1 and 3).} \\
  \multicolumn{5}{l}{\commentb  $C= 2457996.842150 + 0.02535754750 E$.} \\
  \multicolumn{5}{l}{\commentc Number of points used to determine the maximum.} \\
\end{tabular}
\end{center}
\end{table*}

\setcounter{table}{3}
\begin{table*}
\caption{Timings of superhump maxima of NSV 1440 in 2017 (continued).}
\begin{center}

\begin{tabular}{ccccc}
\hline
$E$ & Maximum time\commenta & Error & $O-C$\commentb & $N$\commentc\\
\hline
229 & 12.84706 & 0.00050 & -0.00319 & 16 \\
230 & 12.87103 & 0.00067 & -0.00458 & 16 \\
231 & 12.89907 & 0.00047 & -0.00190 & 13 \\
266 & 13.78420 & 0.00084 & -0.00428 & 19 \\
267 & 13.81098 & 0.00090 & -0.00286 & 16 \\
268 & 13.83904 & 0.00237 & -0.00015 & 15 \\
269 & 13.86463 & 0.00101 & 0.00008 & 16 \\
270 & 13.88976 & 0.00113 & -0.00015 & 15 \\
346 & 15.80863 & 0.00226 & -0.00846 & 12 \\
347 & 15.83535 & 0.00155 & -0.00709 & 12 \\
724 & 25.40132 & 0.00057 & -0.00092 & 59 \\
725 & 25.42540 & 0.00069 & -0.00219 & 59 \\
726 & 25.44882 & 0.00167 & -0.00413 & 57 \\
727 & 25.47992 & 0.00157 & 0.00161 & 59 \\
728 & 25.50450 & 0.00089 & 0.00083 & 58 \\
729 & 25.52809 & 0.00099 & -0.00094 & 59 \\
730 & 25.55569 & 0.00116 & 0.00130 & 58 \\
731 & 25.58045 & 0.00084 & 0.00071 & 58 \\
739 & 25.78447 & 0.00196 & 0.00187 & 10 \\
740 & 25.80980 & 0.00653 & 0.00184 & 9 \\
741 & 25.83844 & 0.00077 & 0.00513 & 12 \\
857 & 28.78673 & 0.00060 & 0.01194 & 10 \\
858 & 28.81273 & 0.00085 & 0.01258 & 10 \\
859 & 28.83715 & 0.00485 & 0.01164 & 13 \\
860 & 28.86360 & 0.00104 & 0.01274 & 14 \\
897 & 29.80434 & 0.00150 & 0.01525 & 10 \\
898 & 29.82538 & 0.00281 & 0.01093 & 13 \\
899 & 29.85278 & 0.00144 & 0.01297 & 13 \\
\hline
\end{tabular}
\end{center}
\end{table*}

\end{document}